\documentclass[prd, twocolumn, nofootinbib, floatfix]{revtex4}

\usepackage{epsfig} 
\usepackage{amsmath}
\usepackage{amsbsy,color}

\newcommand{\beq}{\begin{equation}}
\newcommand{\eeq}{\end{equation}}
\newcommand{\beqa}{\begin{eqnarray}}
\newcommand{\eeqa}{\end{eqnarray}}
\newcommand{\om}{\Omega_m}
\newcommand{\omw}{\Omega_w}

\newcommand{\wtot}{w_{\rm tot}}
\newcommand{\nacc}{N_{\rm acc}}
\newcommand{\ndec}{N_{\rm dec}}
\newcommand{\lcdm}{\Lambda{\rm CDM}}

\begin{document} 

\title{Uniqueness of Current Cosmic Acceleration}
\author{Eric V.\ Linder} 
\affiliation{Berkeley Lab \& University of California, Berkeley, CA 94720, USA \\ 
Institute for the Early Universe, Ewha Womans University, Seoul, Korea} 
\date{\today}

\begin{abstract} 
One of the strongest arguments against the cosmological constant as an 
explanation of the current epoch of accelerated cosmic expansion is the 
existence of an earlier, dynamical acceleration, i.e.\ inflation.  We 
examine the likelihood that acceleration is an occasional phenomenon, 
putting stringent limits on the length of any accelerating epoch between 
recombination and the recent acceleration; such an epoch must last less 
than 0.05 e-fold (at $z>2$) or the matter power spectrum is modified 
by more than 20\%.  
\end{abstract} 

\maketitle

\section{Introduction \label{sec:intro}}

Cosmic acceleration holds the key to physics beyond the standard model 
of particle physics and gravitation.  One of the great puzzles is why 
its characteristic energy scale is so much less than other energy scales 
in the standard model.  No physical principle seems to explain the finite 
magnitude required for the cosmological constant.  As well, one could ask 
what is so special about today that acceleration is just now coming to 
dominate the expansion of the universe.  This is even more of a telling 
argument against a cosmological constant explanation in that its lack of 
dynamics adds a coincidence problem to the fine tuning.  Furthermore, 
another epoch of accelerated expansion is known in the early universe -- 
inflation -- and this was decidedly dynamical and not a cosmological 
constant since it ended. 

This leads one to ask: if we have two periods of acceleration, why not more? 
Could acceleration be an occasional phenomenon?  Such a view ameliorates 
the coincidence problem, since the present is one of many epochs where 
such physics is manifest.  Various models have been proposed to achieve 
this, e.g.\ a high energy physics ``slinky'' potential \cite{barenboim} and 
stochastic beating between multiple fields \cite{dodelson}.  This has 
also been addressed phenomenologically, notably by \cite{griest}, and 
oscillating field models that can achieve this are common in the 
literature (e.g.\ see \cite{mangano,feng,linosc} and references therein).  
Note that even such a technically natural and well motivated model such 
as a pseudo-Nambu Goldstone boson \cite{frie95} in fact goes through 
numerous cycles of acceleration (see Fig.~3 of \cite{calde}).  Any scalar 
field oscillating about a potential minimum will possess equation of state 
$w=-1$ at the turning points where the kinetic energy is zero, and so 
can cause acceleration if its energy density is sufficiently large. 

So the uniqueness of current acceleration is a question of key interest 
not just from the perspective of the cosmological constant and the 
coincidence problem, but also as a guide to the type of physics behind 
cosmic acceleration.  In Sec.~\ref{sec:acc} we examine the general 
characteristics of and constraints on acceleration pre- and 
post-recombination.  In Sec.~\ref{sec:grow} we investigate the effects 
of persistent acceleration post-recombination on the matter density, growth 
of structure, and distances, placing constraints on the length and onset of 
any such period for the two general, comprehensive scenarios.

\section{Persistence of Acceleration \label{sec:acc}}

\subsection{Very Early Universe} 

Acceleration in the early universe is difficult to probe, except 
at particular epochs or if it lasts for many e-folds of expansion. 
Above 1 TeV in energy, such acceleration falls under the rubric of 
inflation, and we do not have tight constraints on the energy/time 
scale or number of individual periods.  Once dark matter is detected 
and understood, we may be able to use the freezeout abundance, involving 
the competition between the interaction rate and the dilution rate due 
to expansion, to probe the cosmic expansion in the 1 GeV -- 1 TeV region. 

Around the time of primordial 
nucleosynthesis, when the energy scale was $\sim1$ MeV, the expansion 
rate is much better known \cite{kaplinghat}.  Radiation dominates, 
with total equation of state $\wtot=+1/3$.  The {\it evolution\/} in 
the expansion rate, i.e.\ the Hubble parameter $H$ as a function of 
scale factor $a$, is less well determined, 
as seen in Fig.~6 of \cite{kaplinghat} (also see \cite{masso,dutta}).  
Recalling that 
\beq 
\frac{d\ln H^2}{d\ln a} = -3\,(1+\wtot)\,, 
\eeq 
we see that the total equation of state (with $\wtot <-1/3$ determining 
acceleration) is not well known even 
during this well understood and tested period.  We can translate 
the variation in the slopes of the expansion behaviors in Fig.~6 of 
\cite{kaplinghat} into the constraint $-0.4<\wtot<1.4$ for their 
conservative case and $0<\wtot<1$ for their tighter case. 

The recombination and decoupling epochs of the cosmic microwave 
background (CMB) are also well measured.  The decoupling depends 
again on the competition between the ionization rate and the 
expansion rate.  The explicit dependence on the function $H(a)$ was described 
analytically by \cite{lin98}, generalizing \cite{joneswyse}.  Again, 
the magnitude $H(a)$ is better determined \cite{zahn} than the logarithmic 
derivative $d\ln H^2/d\ln a$ that gives information on the total 
equation of state.  Acceleration during and pre-recombination will 
be treated in future work.

\subsection{Post-Recombination Acceleration} 

We concentrate on possibilities for acceleration between 
decoupling and the onset of the current acceleration, at redshifts 
$z\approx1-1000$.  Most of this period is not well probed by distance 
measures (but see the next section) so we do not have direct measures 
of the expansion.  However, acceleration breaks matter domination, 
diluting the matter density and 
affecting the growth of structure in two ways.  First, it reduces the 
source term for matter perturbation growth, and second it increases the 
Hubble friction term, proportional to $3-q$ where $q\equiv -a\ddot a/\dot a^2$ 
is the deceleration parameter.  In terms of the total equation of state, 
the friction term is proportional to $5-3\wtot$. 
See Eq.~(8) of \cite{expgro} and Sec.~4.1 of \cite{linrop} 
for more details.  

The effects of intermediate acceleration on growth were discussed for 
certain models in detail by \cite{darkages}.  That article found two 
key effects on growth: the suppression during the actual period of 
acceleration, but also the stunting of the growth {\it rate\/}, i.e.\ 
the velocity $\dot\delta$ of the perturbations, that persisted to much 
later times.  We continue the exploration here, allowing more degrees 
of freedom to the model for acceleration and looking for analytic 
scalings of the effects, while giving a more general and systematic 
treatment. 

Consider the total equation of state of the universe, 
\beq 
\wtot=w\,\Omega_w(a) \label{eq:wtot}
\eeq 
for a flat universe.  To attain acceleration a necessary but not sufficient 
condition is $w<-1/3$; one requires $\wtot<-1/3$.  
Equation~(\ref{eq:wtot}) immediately suggests two alternatives for 
achieving $\wtot<-1/3$.  The first is to have a large dark energy 
fraction $\Omega_w(a)$, hence dark energy will dominate over matter. 
However, once the dark energy dominates it will continue to do so while 
$w<0$ and we do not achieve the desired occasional, or stochastic, nature 
of the accelerating epoch.  
Thus in this approach we will need to allow $w>0$ at some time to undo 
the dark energy domination.  We refer to this as the ``superdecelerating'' 
scenario.  

The second approach is to keep $\Omega_w(a)<\Omega_m(a)$ 
but attain acceleration through a strongly negative dark energy equation 
of state.  Thus the universe accelerates but dark energy does not 
dominate -- this is in fact exactly what happens in $\Lambda$CDM in 
the range $z\approx0.4$--0.7.  Then there is no need for a subsequent 
period that reduces the dark energy density fraction, hence no need 
for $w>0$.  We will see that in order to get an early acceleration period 
separate from the current acceleration, this requires $w\ll-1$.  We 
therefore refer to this alternative as the ``superaccelerating scenario''. 

\begin{figure}
  \begin{center}{
  \includegraphics[width=\columnwidth]{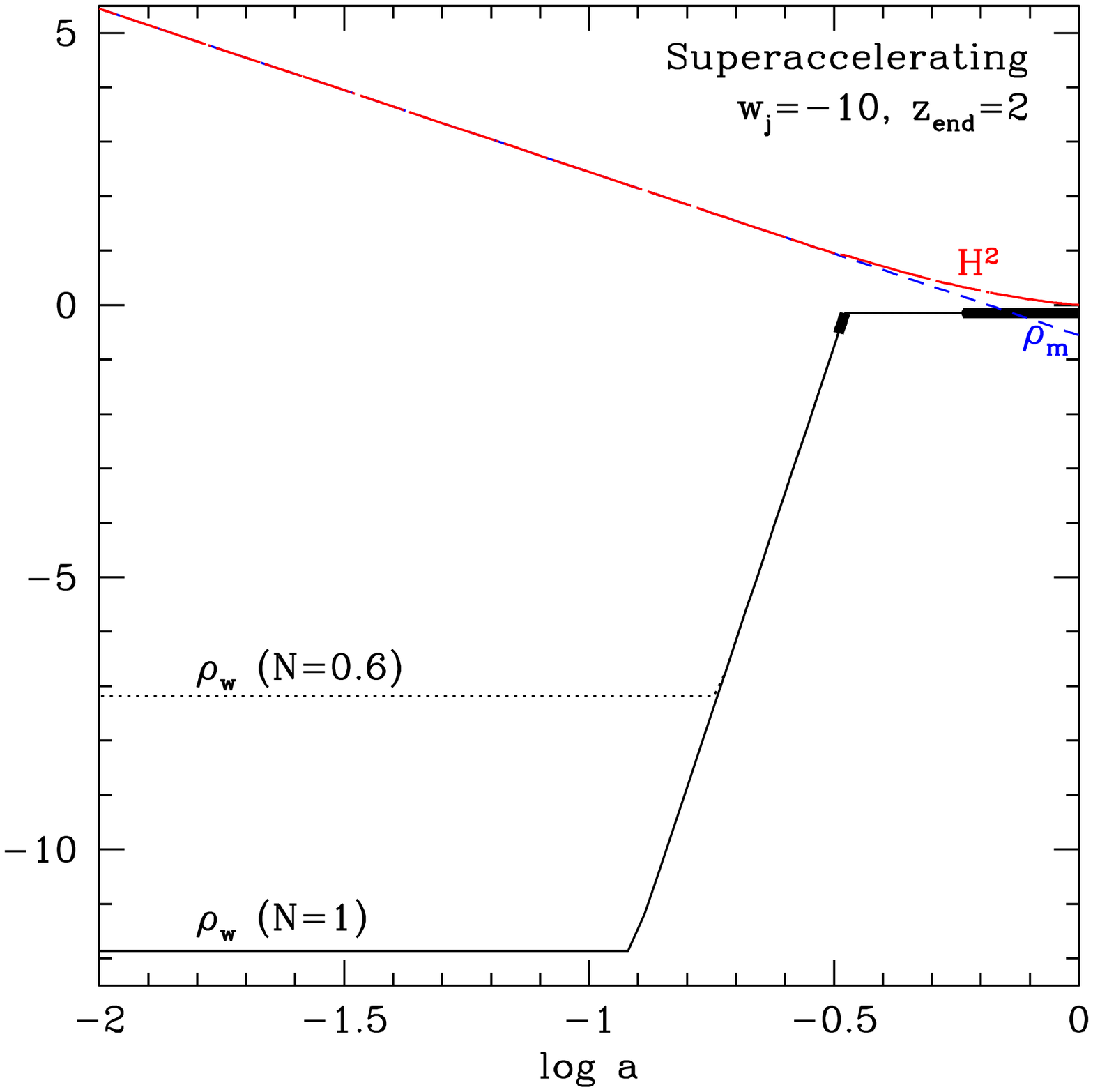}\\ 
  \includegraphics[width=\columnwidth]{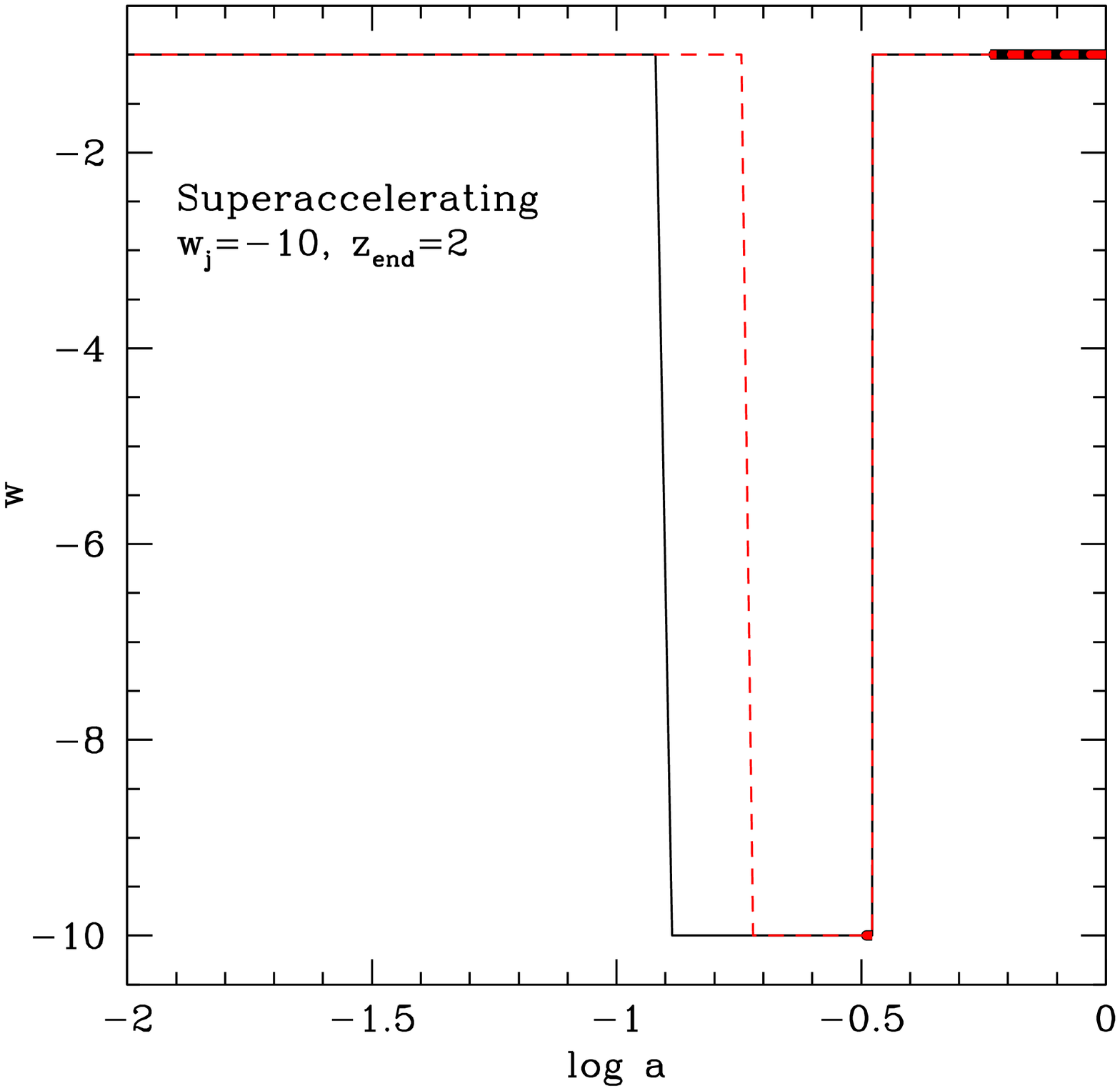}
  }
  \end{center}
  \caption{The superaccelerating scenario has a period of 
strongly negative dark energy equation of state $w_j<-1$, which drives 
$\wtot<-1/3$ and so causes acceleration.  The top panel plots the 
logs of the dark energy density $8\pi G\rho_w(a)/(3H_0^2)$, 
dark matter density $8\pi G\rho_m(a)/(3H_0^2)$, and Hubble parameter 
$H^2/H_0^2$ vs.\ the log of the scale factor $a$.  The $w_j<-1$ epoch 
is taken to last for $N=0.6$ or 1 e-folds, ending at $z_u=2$. 
However, the 
actual acceleration lasts for a much shorter time, shown by the thick 
portions of the $\rho_w$ curves (including for the current epoch of 
acceleration).  In particular, note the early acceleration only lasts 
for 0.034 e-folds. 
The bottom panel plots $w(a)$ on the same horizontal scale so one 
can see directly the effect of $w(a)$ on the densities.  The accelerated 
epochs are again indicated by the thick parts of the curves. 
}
  \label{fig:rhowa}
\end{figure}

\begin{figure}[!ht]
  \begin{center}{
  \includegraphics[width=\columnwidth]{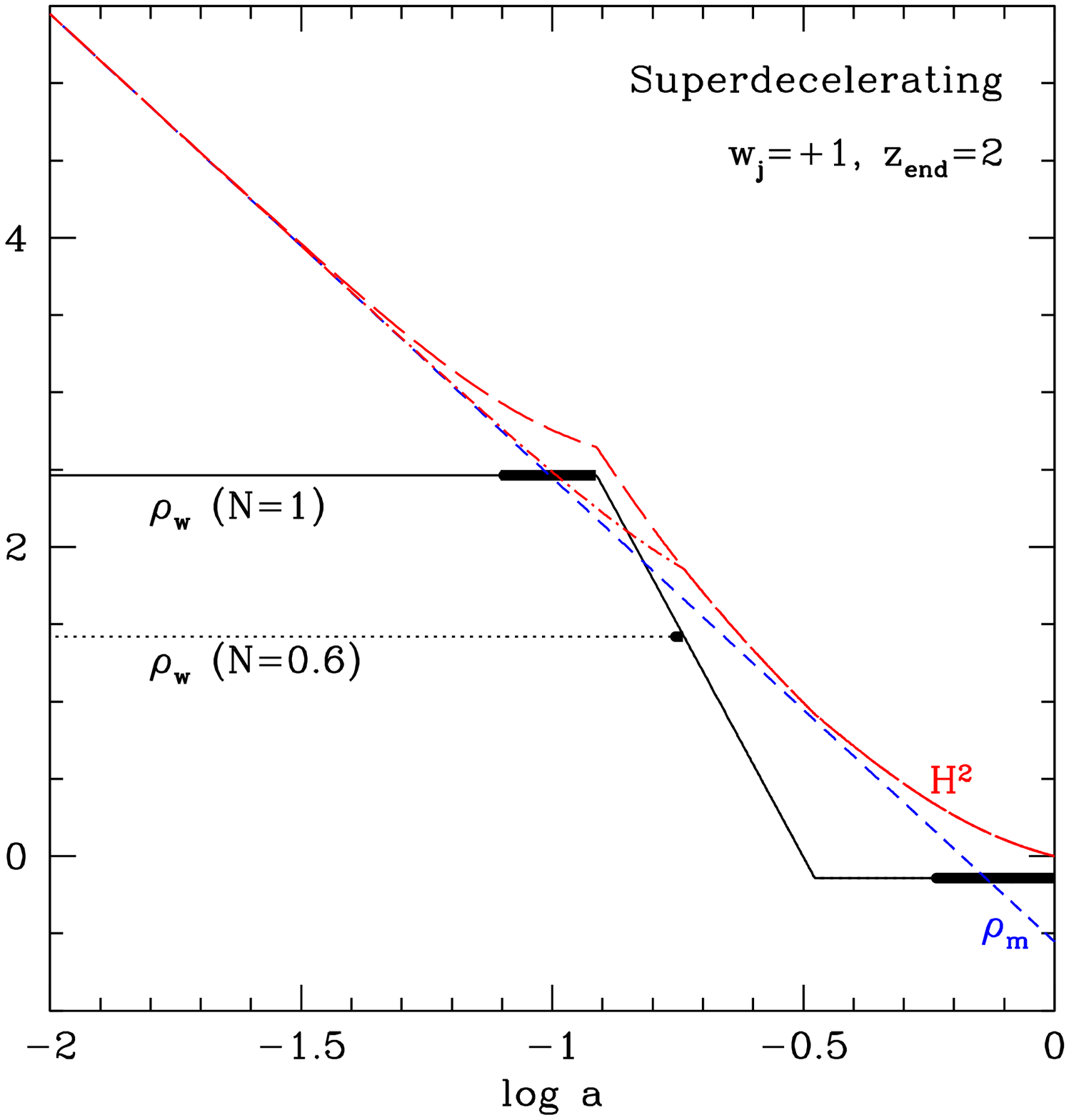}\\ 
  \includegraphics[width=\columnwidth]{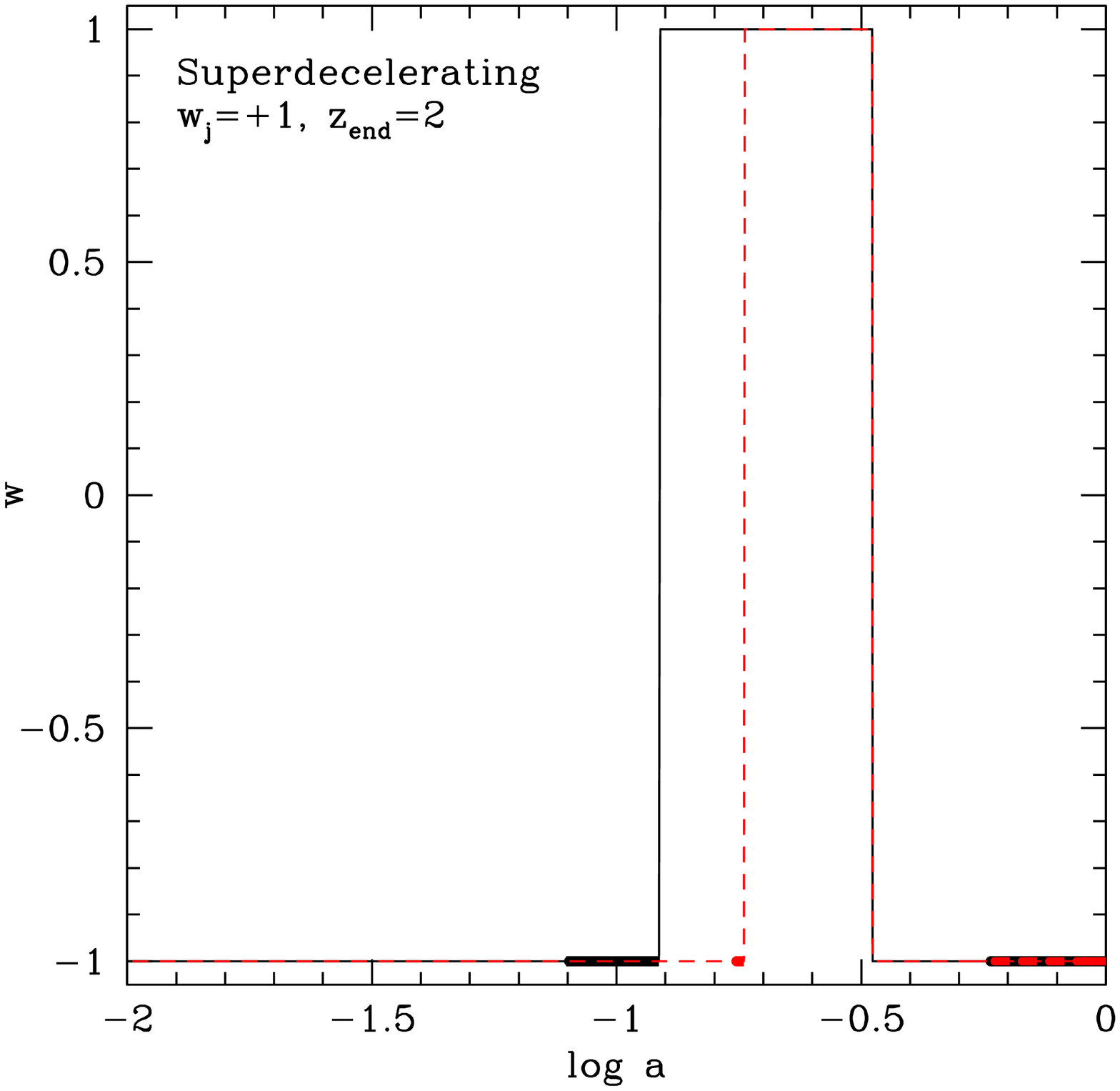}
  }
  \end{center}
  \caption{As Fig.~\ref{fig:rhowa}, but for the superdecelerating 
scenario.  Here the acceleration occurs because of relatively large 
$\rho_w(a)$, rather than supernegative $w$.  To bring the dark 
energy density to the current level then requires a period of 
superdeceleration, where $w>0$ to redshift away the excess density.  
Note that in the $N=0.6$ case the dark energy never dominates the 
matter density during the early acceleration. 
}
  \label{fig:rhowd}
\end{figure}

Figures~\ref{fig:rhowa}-\ref{fig:rhowd} illustrate the characteristics 
of the two different scenarios.  Note that the superacceleration case has 
a lower value of the dark energy density in the past than today 
(effectively a smaller cosmological constant), while the superdeceleration 
case has a higher value.  We show the behaviors of the matter density 
$\rho_m(a)$, dark energy density $\rho_w(a)$, and the Hubble parameter 
$H^2(a)$, as well as the dark energy equation of state $w(a)$.  For the 
last quantity we take a simple model of a constant deviation down (up) 
from the cosmological constant value $w=-1$ for superacceleration 
(superdeceleration), for $N$ e-folds of expansion.  Note that the 
number of e-folds of early acceleration, $N_{\rm acc}$ is much smaller 
than the duration $N$ of the deviation in $w$.  The periods of acceleration 
are shown by the thick parts of the $\rho_w$ or $w$ curves.

\section{Paths to Acceleration \label{sec:grow}} 

Let us examine the scenarios in detail.  These two possibilities of 
superacceleration and superdeceleration span the range of ways to 
achieve acceleration, stemming from the simple definition $\wtot<-1/3$. 

\subsection{Superaccelerating Scenario}

In the superacceleration 
scenario one has a down-then-up transition in the dark energy 
equation of state.  We model this as a simple step away from 
a constant $w$, to a constant $w_j$, lasting for a number of 
e-folds $N=\Delta\ln a$ and then returning to $w$ at scale factor 
$a_u=1/(1+z_u)$ (so the overall deviation begins at $a_d=a_u e^{-N}$).  
We usually take $w=-1$ so that before and after the deviation the 
universe follows $\Lambda$CDM; in any case it is 
matter dominated at high redshift and has a present epoch of acceleration.  
All physical quantities such as densities are continuous, while if we 
had made steps in $\rho$ or $H$ then we would have faced infinities in 
the equation of state.  Although 
there is no problem with our sharp steps, one could also use a smoothed 
form for $w(a)$ capable of rapid transitions, such as the e-fold model 
\cite{linhut05}.  Such formal smoothing makes no difference to the results. 

The dark energy density never dominates until the usual time near the 
present, but nevertheless a highly negative equation of state can drive 
accelerated expansion.  
The condition for accelerated expansion is determined by the total 
equation of state: 
\beqa 
\wtot&=&w\Omega_w(a_d<a<a_u)=w_j\, 
\left[1+\frac{\om}{\omw}a_u^{-3(w_j-w)} a^{3w_j}\right]^{-1} \nonumber\\
&<& -\frac{1}{3}\,. 
\eeqa 
This imposes the requirement that to obtain {\it any\/} acceleration 
one needs 
\beq 
w_j<-\frac{1}{3}\left(1+\frac{\om}{\omw}\,a_u^{3w}\right)\,. 
\eeq 
If we want the early acceleration to be distinct from the current 
acceleration, then we could impose $z_u>2$, say, implying $w_j<-3.8$ 
(taking $\om=0.28$). 
Thus we indeed require a highly negative equation of state. 

\begin{figure}
  \begin{center}{
  \includegraphics[width=\columnwidth]{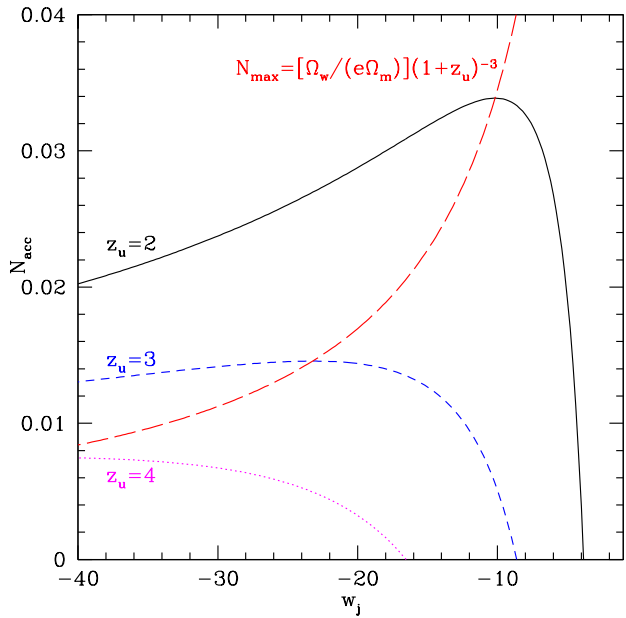}
  }
  \end{center}
  \caption{The superaccelerating scenario can only support a 
very small number of e-folds of acceleration because for a 
supernegative equation of state $w_j$ the dark energy density must start 
very small to end up at the observed level today.  The plot shows 
the number of e-folds of acceleration as a function of $w_j$ and 
the redshift $z_u$ when superacceleration ends.  The red, long-dashed 
curve show the loci of the maximum number of e-folds for each choice 
of $z_u$ and its equivalent $w_j(N_{\rm max})$.  For an early period 
of acceleration sufficiently distinct from the present acceleration, 
e.g.\ $z_u<2$, less than 0.034 e-folds of acceleration are allowed. 
}                                                                              
\label{fig:nacc}                                                             
\end{figure}

In general the number of e-folds of accelerating expansion is given by 
\beq 
N_{\rm acc}=\frac{1}{-3w_j}\,\ln\left[\frac{\omw}{\om} a_u^{-3w}\,(-3w_j-1)\right]\,. \label{eq:naccwj} 
\eeq 
This is plotted in Fig.~\ref{fig:nacc} as a function of $w_j$ for 
various values of $z_u$.  (Note that it is actually independent of the 
number of e-folds the $w_j$ transition lasts, as long as $N>N_{\rm acc}$.)   
The maximum number of e-folds allowed for a 
given $z_u$ is 
\beq 
N_{\rm acc,max}=\frac{1}{-3w_{j,{\rm max}}-1}\approx \frac{\omw}{e\,\om}\, 
a_u^{-3w}\,, 
\eeq 
where $w_{j,{\rm max}}$ is the value of $w_j$ that maximizes 
Eq.~(\ref{eq:naccwj}) for a fixed $a_u$.  
For example, for $z_u=2$ and $w=-1$, one has $w_{j,{\rm max}}=-10.3$ and 
$N_{\rm acc,max}=0.034$.  This relation is shown in Fig.~\ref{fig:nacc} 
as the red, long-dashed curve.

Increasing the redshift of the early acceleration, to alleviate the 
coincidence problem more convincingly, decreases the period of 
acceleration allowed.  
Because such a highly negative equation of state is required, this 
means that the dark energy density grows rapidly, 
$\rho_w\sim a^{-3(1+w_j)}$, so to achieve the same $\Omega_w$ today 
requires a small $\rho_w$ at early times.  
Moving the acceleration period to earlier times reduces 
$\omw(a)$ for the same $\rho_w(a)$, and 
this in turn makes it harder to 
achieve $\wtot=w\Omega_w(a)<-1/3$ at early times.  Attempting to drive 
$w_j$ even more negative simply diminishes early $\Omega_w(a)$ even more, 
and so this scenario can never achieve long periods of early acceleration. 

In summary, at best far less than a single e-fold, only 
$N\approx0.03$, of early acceleration can be achieved.  
Thus one seems to replace a coincidence problem with a further fine tuning 
issue: why should acceleration last for such an apparently unnaturally 
short time, much less than one e-fold?  Inflation lasted for some 60 
e-folds and current acceleration appears to have existed for at least 
half an e-fold so far. 

Since the number of early acceleration e-folds is so small, 
it should have negligible effect on growth 
or distance probes.  For example, even $N=1$ e-fold of $w_j=-10$, ending 
at $z_u=2$, changes the total growth to the present by only 1.7\% 
relative to the standard $\lcdm$ case.  Oddly, growth is actually 
increased by such an acceleration scenario since the diminished 
dark energy density for much of the history slightly increases $\om(a)$. 

The velocity effect on the growth, i.e.\ the stopping and restarting 
of the growth due to acceleration and restoration to deceleration, can 
be measured via the Linder-White \cite{linwhite} matching prescription 
for the nonlinear matter power spectrum (cf.\ \cite{darkages}).  Here, 
the key parameter is the growth ratio $R_g=g(a=0.35)/g_0$ where 
$g(a)=D(a)/a$, $g_0\equiv g(a=1)$, and $D(a)\sim\delta(a)$ is the 
matter perturbation growth factor; this is only suppressed by 0.1\%, 
due to the extremely short duration of the acceleration. 
The distance to CMB last scattering is increased by 0.4\%, essentially 
totally from the actual acceleration period, not the transition to 
$w_j$ per se.

\subsection{Superdecelerating Scenario} 

If, unlike in the superaccelerating scenario, the dark energy density 
dominates during the early acceleration, then 
this must be undone to restore matter domination allowing growth of 
structure.  That is, there must be an epoch after acceleration in 
which the dark energy redshifts away more swiftly than the matter 
so the matter can rise (again) to dominance.  This requires a period when 
$w>0$, which we call superdeceleration.  

In order to have early 
acceleration, one will need to have a higher than standard dark energy 
density at high redshift, in order to obtain the same dark energy density 
today despite the extra dilution (see Fig.~\ref{fig:rhowd}).  In fact, 
the dark energy does not even have to dominate, only have $\omw(a)>1/(-3w)$, 
for example $\omw(a)>1/3$ for $w=-1$, to give rise to acceleration.  
We take a model with $\lcdm$ at 
high redshift, though with extra dark energy density.  This high value 
of dark energy density kicks off a period of early acceleration, 
lasting for $\nacc$ e-folds.  Then, at $z_j$ the dark energy equation 
of state jumps up to $w_j>0$ (note the jump is up not down, in contrast 
to the superacceleration scenario).  This superdecelerating phase lasts 
for $\ndec$ e-folds, then $w=-1$ is restored and the expansion proceeds 
again as $\lcdm$. 

\begin{figure}
  \begin{center}{
  \includegraphics[width=\columnwidth]{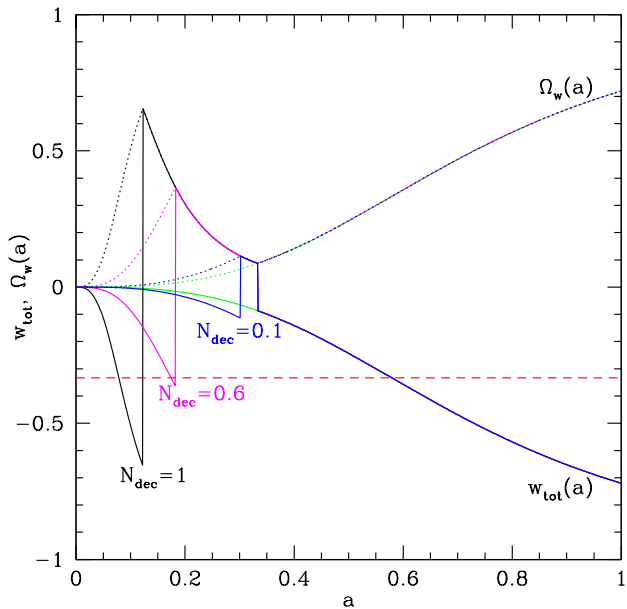}
  }
  \end{center}
  \caption{The fractional dark energy density $\omw(a)$ (dotted, 
positive only curves) 
and total equation of state $\wtot(a)$ (solid curves) are plotted for 
the superdecelerating scenario with three different values of $\ndec$.  
Note that the $\ndec=0.1$ case never achieves early acceleration 
($\wtot<-1/3$, shown by the horizontal dashed line), the $\ndec=0.6$ case 
barely does, and even the $\ndec=1$ model has only a short period of 
early acceleration.  All models shown have $w_j=+1$ and return at $z=2$ 
to $\lcdm$ with $\Omega_w(a=1)=0.72$.  The smooth, 
green curves correspond to $\lcdm$ without any transition. 
}
  \label{fig:omdec}
\end{figure}

Figure~\ref{fig:omdec} shows the behavior of the fractional dark energy 
density $\omw(a)$ and total equation of state $\wtot$.  The recent 
universe is standard $\lcdm$, while there are epochs of acceleration 
($\wtot<-1/3$) and superdeceleration ($\wtot>0$) in the early universe.  
The dark energy density 
is boosted at high redshift relative to the standard case.

One sees that a substantial period of superdeceleration is required 
to permit even a brief period of early acceleration.  Indeed, no 
early acceleration can occur unless 
\beq 
\ndec>\frac{1}{3(1+w_j)}\,\ln\left[\frac{\om}{2\omw a_j^3}\right] 
\,. \label{eq:nacc0} 
\eeq 
For $z_j=4$, for example, one requires $\ndec>0.53$.  In general, the 
number of e-folds of early acceleration, $\nacc$, is related to $\ndec$ 
and the transition scale factor from acceleration to deceleration, $a_j$, 
by 
\beq 
\nacc=\ndec\,\left(1-\frac{w_j}{w}\right)+\ln a_j-\frac{1}{3w} 
\ln \frac{2\omw}{\om}\,, 
\eeq 
where $w$ is the dark energy equation of state outside the superdeceleration 
period (i.e.\ $w=-1$ if we take $\lcdm$ as the baseline). 

\begin{figure}
  \begin{center}{
  \includegraphics[width=\columnwidth]{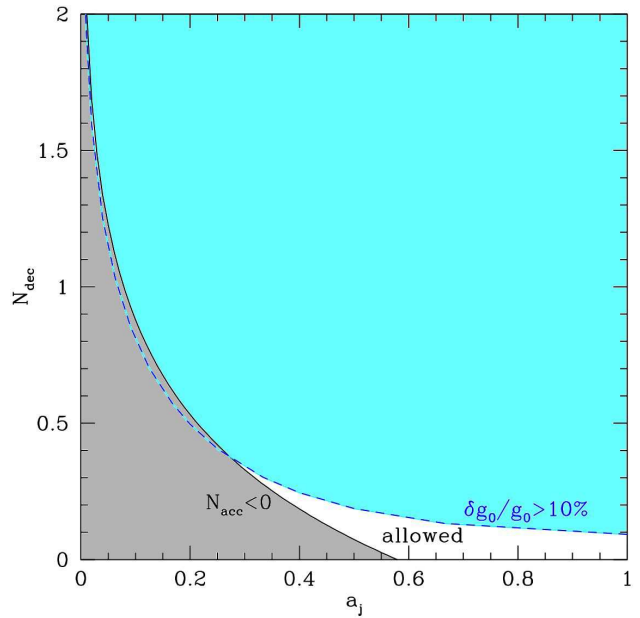}
  }
  \end{center}
  \caption{Early acceleration due to a high early dark energy density 
can only occur, and still deliver the same dark energy density today, 
if it is followed by sufficient superdeceleration (here $w_j=+1$).  
Only the space above 
the solid, black curve, giving $\nacc>0$, achieves this.  Too much 
superdeceleration, however, strongly suppresses the total growth to 
the present.  Only the space below the dashed, blue curve diminishes 
the growth relative to $\lcdm$ by less than 10\%.  The only surviving 
region is the thin crescent between the curves at $a_j>0.27$, or 
$z_j<2.7$, which hardly qualifies as early acceleration.
}
  \label{fig:gsuperd}
\end{figure}

Equation~(\ref{eq:nacc0}) gives a lower limit on $\ndec$ in terms of 
$a_j$.  If $\ndec$ is too large, however, then growth will be severely 
affected.  Recall that any deviation from matter domination, whether 
acceleration or superdeceleration, can suppress growth.  This constraint 
then gives an upper limit on $\ndec$ so we can evaluate the viability of 
the superdecelerating scenario for early acceleration based on the 
comparison of these upper and lower limits.  From 
Fig.~\ref{fig:gsuperd} we see that in fact superdeceleration fails.

Only a thin sliver of parameter space for $z_j<2.7$ survives, and 
such a late epoch of acceleration does nothing to help the coincidence 
problem (decreasing $w_j<+1$ worsens the situation).  In addition, the 
length of the acceleration is quite short: 
for $z_j=1.5$, say, one has at best a number of e-folds $\nacc=0.12$.  
This does not seem 
like a very natural value, being much less than an e-fold.  Attempting 
to create several epochs of acceleration, necessitating even shorter 
durations, simply exacerbates the problems.  At high redshift, the 
curve $\nacc=0$ roughly corresponds to $\delta g_0/g_0=15\%$ 
(one can think of this roughly as a shift in the mass amplitude $\sigma_8$), 
or 30\% deviation in the matter power spectrum amplitude.  

The growth velocity factor, measured through $R_g$, is affected at the 
5.1\% (2.7\%) level for $z_j=1.5$ (2.0) and the maximum number of 
e-folds of acceleration, 0.12 and 0.05 respectively.  More severe is 
the impact on the distance to last scattering, mostly due to the 
higher early dark energy density: this is reduced by 2.9\% (2.7\%), 
which essentially means that even the thin sliver of allowed parameter 
space in Fig.~\ref{fig:gsuperd} is in doubt when including CMB constraints. 
Compensating by decreasing the matter density can preserve the distance 
to last scattering, but worsens the growth deviation.

\section{Conclusions \label{sec:concl}} 

Only two paths exist to obtaining a period of cosmic acceleration: 
the dark energy density is subdominant but its equation of state is 
highly negative (superacceleration), or the dark energy density is 
dominant, or nearly so, and its equation of state is at least moderately 
negative (superdeceleration).  In the first case, we have seen that 
the dynamics unavoidably forces the dark energy density to be so low 
that acceleration is quite transient -- less than 0.035 e-folds for 
acceleration before $z=2$.  This conclusion only depends on a dark energy 
density fraction today $\Omega_w\approx 0.7$, not on any external data. 

In the second case, the acceleration is caused by an unusually 
{\it high\/} early dark energy density, which then must be diluted in order 
to restore matter domination.  To accomplish this, the dark energy equation 
of state must become positive, hence leading to a period of 
superdeceleration.  A period of acceleration requires a longer period of 
superdeceleration, yet these both suppress the growth of matter density 
perturbations.  The two constraints of requiring sufficient superdeceleration 
to allow for some early acceleration, yet not disrupting growth, pinch 
the allowed parameter space to permit at most 0.05 e-folds of acceleration 
before $z=2$. 

Thus, neither possibility for achieving $\wtot<-1/3$ -- via superacceleration 
or superdeceleration -- can truly provide early, or occasional acceleration, 
or provide any help in ameliorating the coincidence problem.  It is 
interesting that such a clear conclusion falls out from such simple 
arguments.  
The current acceleration indeed appears to be unique since 
the time of CMB decoupling.

\acknowledgments

I thank Stephen Appleby, Marina Cort{\^e}s, Roland de Putter, Manoj 
Kaplinghat, and especially Tristan Smith for useful discussions. 
This work has been supported in part by the Director, Office of Science, 
Office of High Energy Physics, of the U.S.\ Department of Energy under 
Contract No.\ DE-AC02-05CH11231, and the World Class University grant 
R32-2009-000-10130-0 through the National Research Foundation, Ministry 
of Education, Science and Technology of Korea.

\end{document}